\documentclass[aps,prb,superscriptaddress,nobibnotes]{revtex4}

\usepackage{graphicx}
\begin{document}

\title{Sharp optical phonon softening close to optimal doping in
  La$_{2-x}$Ba$_x$CuO$_{4+\delta}$}  

\author{Matteo d'Astuto}
\email[]{matteo.dastuto@impmc.jussieu.fr}

\affiliation{Institut de Min\'eralogie et de Physique des
Milieux Condens\'es (IMPMC), Universit\'e Pierre et Marie Curie -
Paris 6, case 115, 4, place Jussieu, 75252 Paris cedex 05, France}
\altaffiliation{Institut de Min\'eralogie et de Physique des Milieux
Condens\'es (IMPMC), CNRS UMR 7590, Campus Boucicaut, 140 rue de
Lourmel, 75015 Paris, France}

\author{Guy Dhalenne}
\affiliation{Laboratoire de Physico-Chimie de l'Etat Solide (ICMMO), CNRS
UMR8182, Universit\'e Paris-Sud 11, B\^atiment 410, 91405 Orsay, France}

\author{Jeff Graf} \affiliation{Materials Sciences Division, Lawrence
  Berkeley National Laboratory, Berkeley, CA 94720, USA}

\author{Moritz Hoesch}
\affiliation{European Synchrotron Radiation Facility,
BP 220, F-38043 Grenoble Cedex, France}

\author{Paola Giura}
\affiliation{Institut de Min\'eralogie et de Physique des
Milieux Condens\'es (IMPMC), Universit\'e Pierre et Marie Curie -
Paris 6, case 115, 4, place Jussieu, 75252 Paris cedex 05, France}

\author{Michael Krisch}
\affiliation{European Synchrotron Radiation Facility,
BP 220, F-38043 Grenoble Cedex, France}

\author{Patrick Berthet}
\affiliation{Laboratoire de Physico-Chimie de l'Etat Solide (ICMMO), CNRS
UMR8182, Universit\'e Paris-Sud 11, B\^atiment 410, 91405 Orsay, France}

\author{Alessandra Lanzara}
\affiliation{Materials Sciences Division, Lawrence Berkeley National
  Laboratory, Berkeley, CA 94720, USA} \affiliation{Department of
  Physics, University of California Berkeley, CA 94720, USA}

\author{Abhay Shukla}

\affiliation{Institut de Min\'eralogie et de Physique des
Milieux Condens\'es (IMPMC), Universit\'e Pierre et Marie Curie -
Paris 6, case 115, 4, place Jussieu, 75252 Paris cedex 05, France}

\date{\today}

\begin{abstract}
We report a direct observation of a sharp Kohn-like anomaly in the doubly 
degenerate copper-oxygen bond-stretching phonon mode occurring at
$\mathbf{q}\mathrm{=(0.3, 0,0)}$ 
in La$_{2-x}$Ba$_x$CuO$_{4+\delta}$ with $\mathrm{x=0.14\pm0.01}$, thanks to  
the high $\mathbf{Q}$ resolution of inelastic x-ray scattering.
This anomaly is clearly seen when the inelastic signal is analysed
using a single mode but is also consistent with a two mode hypothesis
possibly due to a splitting of the degenerate modes due to symmetry
breaking stripes.   
Our observation shows that the effect persists at the stripe
propagation vector in a superconducting system close to optimal
doping. 
\end{abstract}

\pacs{74.25.Kc, 74.72.Dn, 63.20.Dj, 63.20.Kr, 78.70.Ck}


\maketitle

\section{Introduction}
Copper-oxygen bond stretching modes in high-temperature
superconducting cuprate (HTCS) have long been known to show anomalies
related to doping\cite{pintrev1}.   
The origin of the anomalies is still an open issue. Along with
electron-phonon coupling effects, a mechanism involving the
formation of an inhomogeneous charge state, also known as
\textit{stripes}\cite{zaanen,machida,kato,nature-kievelson,rmp-kievelson}
has been discussed \cite{mcqueeney,pinbrief}. 
At the outset let us clarify the anomalies seen in this phonon mode. 
Earlier works\cite{pintrev1} focused on the gradual softening of this
mode with doping, a phenomenon seen quite universally in HTCS and now
relatively well understood, even theoretically \cite{giustino} on the
basis of electron-phonon coupling and screening mechanisms which come
into play on doping.   
This softening can be well described by a cosine-like behaviour of the
mode with its minimum at the zone boundary. 
Even in the earliest works, however, there were indications that the
dispersion might be more complicated, with deviations from the cosine-like
shape \cite{mcqueeney}. 
The possibility that these deviations are caused by stripes has been
invoked and adopted in a recent paper \cite{reznik}
in which previous data on HTCS, are compared to newer ones, in
particular for $\mathrm{La_{2-x}Ba_{x}CuO_{4+\delta}}$ with x=1/8. 
In this system, for x$\sim$1/8, an anomalous suppression of the
superconductivity reported in Ref. \onlinecite{moodenbaugh} was later
shown to be associated with charge and 
spin stripe order\cite{fujita,abbamonte,kim:064520}.   
Signatures of stripes in superconducting
samples remain elusive, though it was proposed that
stripes are difficult to detect in these because they are no longer  
static\cite{vojta:097001,hinkov-nphys}.
Using phonons to probe charge fluctuations can poten-
tially lead to a reliable signature of static and dynamic
stripes in HTCS.
The authors of Ref. \onlinecite{reznik} in describing their high
resolution inelastic neutron scattering data of the copper-oxygen bond
stretching mode, show how one of its two normally degenerate
components follows the expected cosine-like dispersion while the other
deviates presenting a much sharper dip. They interpret this behaviour
as a Kohn-like anomaly due to stripes which lift the degeneracy.
\footnote{The ``classical'' Kohn-anomaly is due to electron phonon
  coupling and in its strongest form can result in a charge density wave. 
The wave vector related to this effect is given by a nesting vector of
the Fermi surface\cite{gruner}. 
In the case of a stripe related anomaly, the difference is that
stripes are supposed to originate from strong on-site coulomb
repulsion\cite{nature-kievelson} and are not linked to the Fermi
surface topology.} 
This scenario is very
intriguing, nevertheless, the dip is not directly visible in the data
of Ref. \onlinecite{reznik}, and the authors deduce its existence
from a broad shoulder on the low energy side of the Cu-O bond
stretching phonon mode. Thus, as we show in a recent paper
\cite{jeff-brief}, it is desirable to have a direct measurement of
the dip since without it, other reasons could be invoked to explain
the observed signal.
To clarify this point we have carried out a high
$(\mathbf{Q},\omega)$-resolution 
measurement in $\mathrm{La_{2-x}Ba_{x}CuO_{4+\delta}}$ 
using Inelastic X-ray Scattering (IXS), with a comparable energy
resolution, but with a higher resolution in reciprocal space compared
to previous inelastic neutron scattering experiments. 

This allowed us not only to confirm the presence of a dip around 
$\mathbf{q}\mathrm{=(0.3, 0,0)}$ in the dispersion of the
copper-oxygen bond stretching mode, but also to extend this observation to
superconducting doping value (x = 0.14).
This doping level leads to a stripe-propagation-vector
$\mathbf{q}\mathrm{\approx(0.3,0,0)}$\cite{rmp-kievelson}, in striking
agreement with our findings, and supports a picture based on the
interaction with stripes \cite{mcqueeney,reznik}.


\section{Experiment}

\begin{figure}
\includegraphics[scale=0.3]{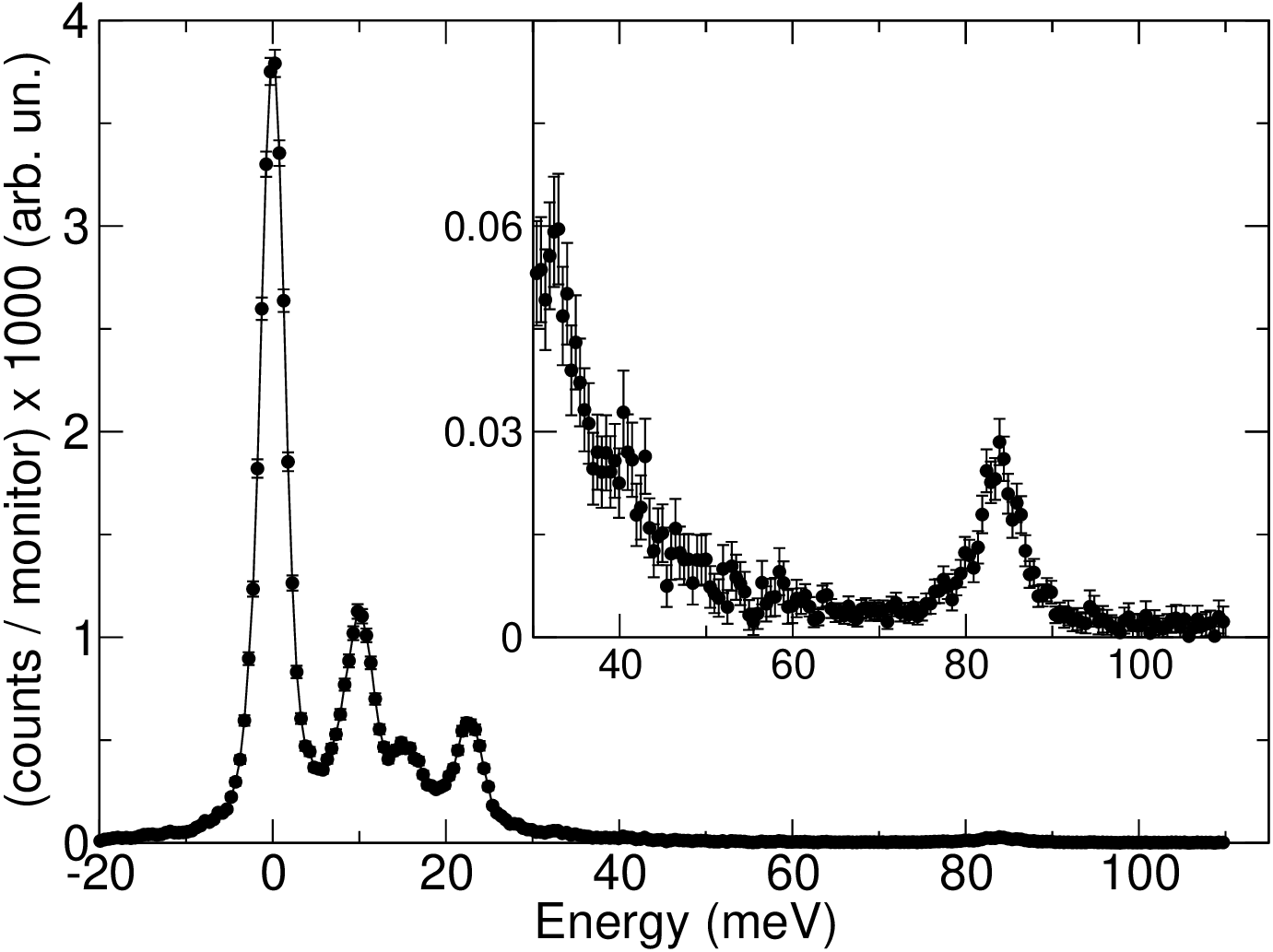}
 \caption{\label{raw-int}
Inelastic x-ray scattering example spectrum of 
$\mathrm{La_{1.86}Ba_{0.14}CuO_{4+\delta}}$, at 
$\mathbf{Q}\mathrm{=(3.11,0.04,0)}$. 
At zero energy one observes the elastic line, while at positive
energy, several phonon modes are observed corresponding to Stokes
processes. 
In the inset we zoom on to the high energy optical Cu-O bond
stretching mode. Anti-Stokes modes are not visible because of the
unfavorable detailed balance at low temperature.}
 \end{figure}

We studied a single crystal of $\mathrm{La_{2-x}Ba_{x}CuO_{4+\delta}}$
grown from the melt in an image furnace by the traveling solvent zone
method, under a pressure of 3 bar oxygen. 
The measured $\mathrm{T_c}$ = 18.1 K with $\Delta\mathrm{T_c}$ = 7 K
defined as the temperature range from $\mathrm{90\%}$ to
$\mathrm{10\%}$ of the maximum Meissner signal, is consistent with
micro-probe results indicating a content of $\mathrm{x=0.14\pm0.01}$
\cite{axe-prl-laba}.     
The experiment was carried out at the IXS beamline II (ID28) at the
European Synchrotron Radiation Facility in Grenoble. 
The sample was mounted in reflection geometry, on the cold finger of a
closed-loop helium cryostat, in order to kept it at constant
temperature $\mathrm{T = (17.8\pm0.5) K}$ during the experiment. 
The low temperature was chosen so as to optimize the signal of the high
energy modes over the tails of low energy ones.  
The sample $c$ crystal axe was perpendicular to the scattering
plane. We consider the tetragonal unit cell with
$\alpha$ = $\beta$ = $\gamma$ = 90$^{\circ}$ and the axes $a$ and
$b$ along the Cu-O bond. The sample was aligned along (H,0,0), and
we refined the parameter $a$=3.792 $\pm$ 0.001 \AA, according to the
$\theta - 2 \theta$ scan on the (4, 0, 0) reflection, and adopted
the value of $c$ = 13.235~\AA. The rocking curve at the (4,0,0)
reflection had a FWHM $\sim ~0.06^{\circ}$, indicating a very
low mosaic spread in the small volume probed. 
The IXS multi-analyzer spectrometer\cite{masciovecchio1,krisch-ixs}
configurations chosen were of 
type $\mathbf{Q}=\mathbf{G} + \mathbf{q} 
\mathrm{= (3 , 0, 0 ) + (}q_x, q_y, 0)$, 
with simultaneous measurements from 7 analyzers. 
We collected data from the first 5 analyzers closer to the beam
direction, which are at fixed angular spacing of $\sim0.75^\circ$
($q_x\sim \pm$ 0.07), with the third one in longitudinal condition
$q_y$ = 0 while for the others $q_y \leq \pm$ 0.045. 
This setup corresponds to the 3rd extended Brillouin Zone, where the
zone center $\Gamma$ is at (4,0,0) and the extended zone boundary at
(3,0,0). 
The standard or folded zone boundary is at the point M=(3.5, 0, 0). 
The scattering vector resolution was set using a slit opening
in front of the analyzers of h $\times$ v = 20 $\times$ 60 mm
corresponding to a solid angle of $\delta\theta \times \delta\xi$ =
0.19$^{\circ}$ $\times$ 0.57$^{\circ}$. 
The resolution in the reciprocal space for 0.6968 \AA ~wave-length
was ($\pm \delta q_x$, 0, 0) with $ \delta q_x \approx $ 0.009.
The high energy resolution
is obtained using the back-scattering silicon monochromator aligned
along the (1 1 1) direction\cite{verbeni,verbeni-rsi}. 
In the present experiment we choose to work with the Si (9 9 9)
reflection order, with a wave-length of 0.6968 \AA (17794 eV) and an
energy resolution $\Delta$E = 3.0 $\pm$ 0.2 meV. 
An example of a raw IXS spectrum is given in Fig. \ref{raw-int}. 
The energy scans were fitted using a sum of
pseudo-Voigt for elastic and resolution-limited inelastic
contributions, with parameters fixed to match the instrumental
function, while a Lorentzian line-shape was used to fit modes with an
intrinsic width larger than the instrumental resolution. 
In order to assign the measured phonon modes, we performed a
lattice dynamical calculation using a computer code \cite{mirone}
based on a shell model\cite{chaplot}. 
More details on analysis and simulation are described elsewhere
\cite{jeff-brief}.

\section{Results and discussion}

\begin{figure}
\includegraphics[scale=0.5]{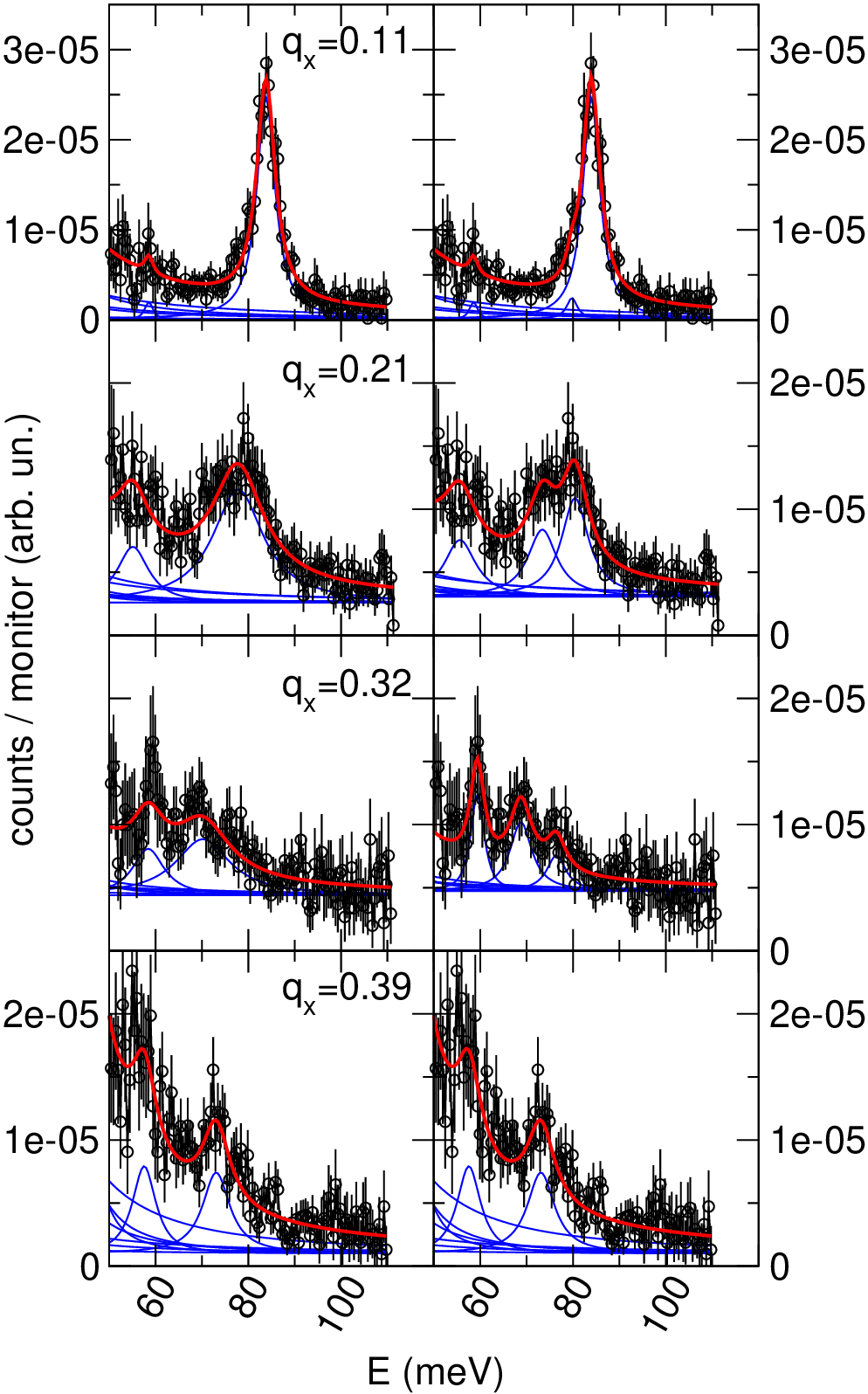}
 \caption{\label{fit}(Color online)
Inelastic X-ray scattering example spectra from 
$\mathrm{La_{1.86}Ba_{0.14}CuO_{4+\delta}}$, at different wave-vectors
$\mathbf{Q}\mathrm{=(3+q_x,q_y,0)}$ (see text). 
In the left column the data are fitted using the degenerate Cu-O bond
stretching mode hypothesis. 
In the right column the data are fitted supposing two Cu-O bond
stretching mode. }
 \end{figure}

The high energy portion of some typical measurements is shown in
Fig. \ref{fit}. In this spectral window, at least two modes could be
identified: a mode at about 60 meV, corresponding to the 5th
longitudinal optic mode (in order from the lowest energy), 
and a second dispersing from about 90 meV down to about 70 meV, the
highest (6th) energy longitudinal Cu-O bond-stretching (or half breathing)
optic mode. The fast downward dispersion is accompanied by a
rapid broadening from 5 meV to about 15 meV FWHM.  
This mode is doubly degenerate, and could eventually split
if the local symmetry is lowered by some anisotropy.  
Such splitting has been suggested \cite{reznik} for a
q$_x\sim$ 0.2 - 0.3, corresponding to the propagation vector of a
charge modulation (``stripe''). 
Other possible origins of the double modes have been previously 
reported\cite{mcqueeney,tranquada-nick} always in connection with
microscopic phase-separation. 
The doubling of the modes can contribute to the observed
broadening if the energy resolution is not enough to distinguish the
two modes as in our case.  
In most of the measured spectra it is then not possible to have
independent fitted parameters for frequency, intensity and width of
each mode. 
We have used the error estimation of the fitted widths as a goodness of 
fit criterion and only in the case of q$_x$ = 0.32 (see Fig. \ref{fit}), 
the two mode fit appears to be better than the single mode fit for our data.

\begin{figure}
\includegraphics[scale=0.35]{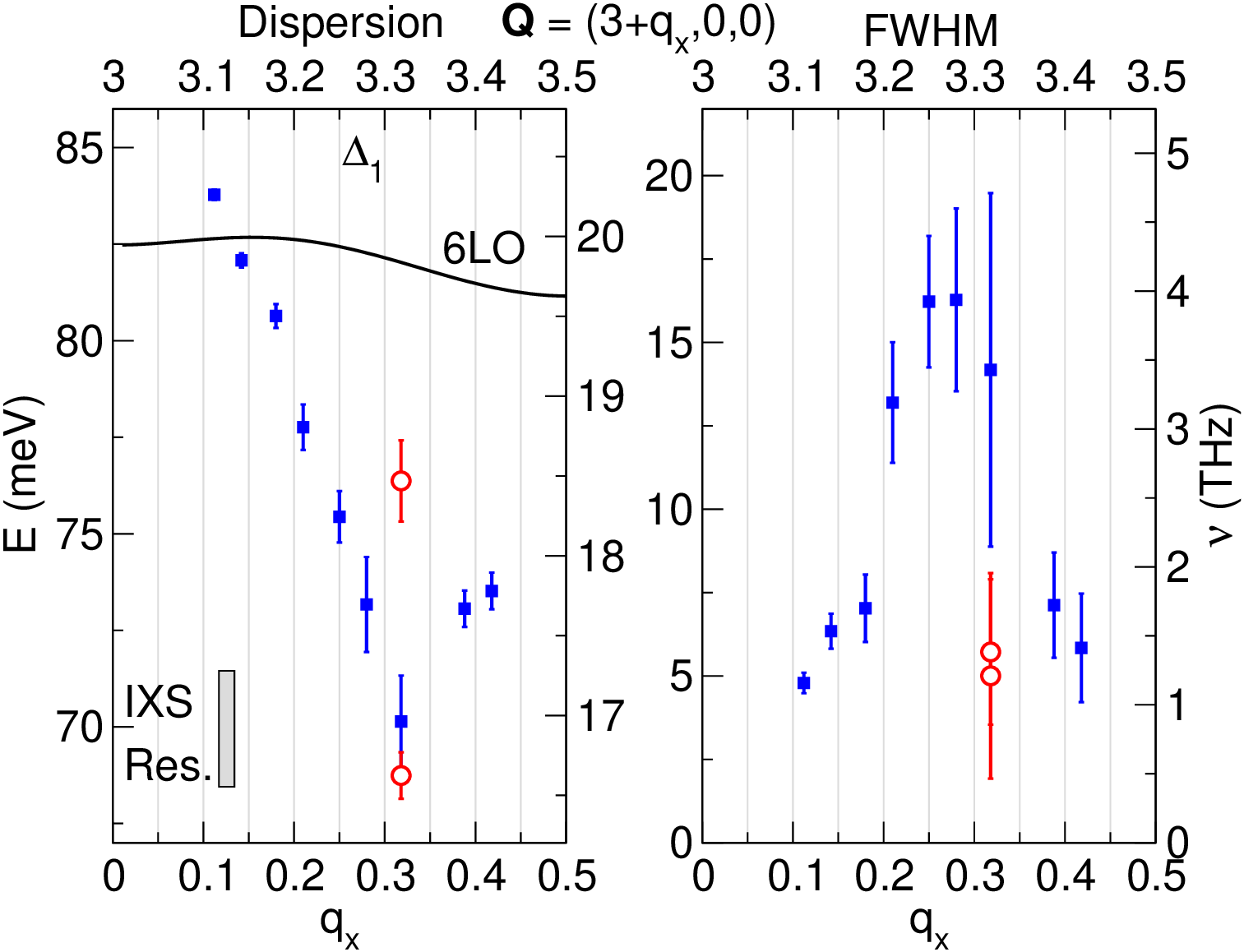}
 \caption{\label{disp} (Color online)
Left panel: 
$\mathrm{La_{1.86}Ba_{0.14}CuO_{4+\delta}}$
Cu-O bond stretching (BS) dispersion. Blue square represent
  frequencies from the model fit with one mode, while red hollow
  circles represent a fit with two modes. The line correspond to
  a shell model lattice dynamics calculation.
Right panel: 
$\mathrm{La_{1.86}Ba_{0.14}CuO_{4+\delta}}$
Cu-O bond stretching (BS) width, symbols as above.
 }
 \end{figure}

In Fig. \ref{disp} we show the results of our fitting for the
frequencies (left panel) and width (right panel) for the single mode
(blue circles). 
A minimum at q$_x$ = 0.32 $\pm$ 0.04 is clearly observed.  
In the same figure, we also show the results for the fit in a double
mode hypothesis (red squares), for q$_x$ = 0.32. 


The difficulty of the analysis (the impossibility of reliably fitting
two closely placed peaks) is inherent to the
experimental strategy chosen: in order to follow in detail the
dispersion we chose a higher $\mathbf{Q}$-resolution with comparable
energy resolution to Ref. \onlinecite{reznik}. This also results in a
significantly different line-shape. 
In Ref. \onlinecite{reznik}, using inelastic neutron scattering, the
highest mode appears more intense because it is weakly dispersing, and
thus benefits from a large $\mathbf{Q}$ integration, while 
the low energy one appears broader because the supposed Kohn-like
dip in the dispersion is only as large as the experimental
$\mathbf{Q}$-resolution.  
Assuming that the two-mode hypothesis is valid let us examine the
consequences for our data. 
A $\mathbf{Q}$-resolution narrower than the dip width results in a
sharper low energy mode. 
We have already seen that the intensity of the high energy mode decreases
due to the better $\mathbf{Q}$-resolution. 
This results in two modes of similar intensity and width, more
difficult to separate with this energy resolution.   
Returning to the results shown in Fig. \ref{disp}, in a one mode
hypothesis, we obtain a maximum of the width around q$_x\sim$ 0.3,
coinciding with the dispersion minimum. 
Indeed, even a one peak fit gives a well pronounced anomalous dip that 
is inconsistent with a cosine-like dispersion.
The dip is observed
even with a one peak fit because the dispersion of the concerned mode
and its intensity are strong enough and dominate even after averaging
with the less dispersing higher energy mode.  
The maximum in the width is, in fact, a measure of the energy distance
between the modes which is associated with the dip, as already pointed
out (see Fig. 4 in Ref. \onlinecite{reznik}). 
Note that we extended the observation of this anomaly to a doping
range which is well beyond the singular 1/8th level and crucially in a 
superconducting range where the stripes should be dynamic.
Consistent with this higher doping level we observe that the
dip is more spread out over the BZ, already starting at q$_x\approx$
0.1 with a FWHM of $\mathrm{4.7\pm0.3}$ meV and that the softening is reduced
even in the two mode hypothesis. 
These observations are thus consistent with a persisting anomaly at the
stripe propagation wave-vector in a more metallic system, where both the
doubling and the shift of the low energy mode appear to be reduced. 
Previous results obtained by IXS \cite{fukuda} on
$\mathrm{La_{2-x}Sr_{x}CuO_{4+\delta}}$ for different doping level are
interpreted in the standard cosine-like dispersion picture, but
corresponds to a set-up with only $\approx$ 6 meV resolution, about
twice the present one. 
A review of this anomaly in other HTCS is reported in Ref. \onlinecite{reznik}. 
Finally, we note here that a possible link has recently been
suggested in single layer
Bi$_2$Sr$_{1.6}$La$_{0.4}$Cu$_2$O$_{6+\delta}$ \cite{jeff-bisco}
between the maximum of the Cu-O  
bond stretching softening and width, and a Fermi surface nesting
vector at $\mathrm{q_x\approx}$0.22, with a clear kink in the electron
dispersion at the energy of the phonon mode. The vector is close to the stripe
propagation vector in double layer Bi$_{2-x}$Sr$_2$CaCu$_2$O$_{8+\delta}$ 
reported by STM in Ref. \onlinecite{Hoffman02} and very recently in 
single layer Bi$_{2-y}$Pb$_{y}$Sr$_{2-z}$La$_{z}$Cu$_2$O$_{6+x}$ \cite{wise}.  
This correlation suggests that stripes are closer 
to the classical charge-density-wave, which in fact fully develops
in others systems as bismuthates, with similar effects on the Bi-O bond
stretching modes \cite{bkbio}. In order to confirm this hypothesis, however,
further investigations, similar to the one of Ref. \onlinecite{jeff-bisco},  
on others systems and dopings are required.

\section{Conclusions}
We directly observe a sharp dip in the 
dispersion of the Cu-O bond-stretching or half-breathing mode at
the wave-vector q$_x\sim$ 0.3 associated with the stripe 
propagation-vector, as previously suggested in Ref. \onlinecite{reznik}.
Our data are consistent with the two mode hypothesis, which
would describe this softening as associated with a doubling of
the mode, possibly due to a splitting of the doubly degenerate Cu-O
bond-stretching mode \cite{reznik}. 
Our observation shows that the effect persists in a superconducting
system close to optimal doping.  

\begin{acknowledgments}
We acknowledge D. Gambetti for technical help.
This work was supported by ESRF through experiment HS-3460 
The IXS measurements and data analysis were
partially supported by the Director, Office of Science, Office of
Basic Energy Sciences, Materials Sciences and Engineering Division, of
the U.S. Department of Energy under Contract No. DE-AC02-05CH11231.  
We acknowledge the support of the National Science Foundation through
Grant No. DMR-0349361 and DMR-0405682, as well as of the University of
California, Berkeley, through France Berkeley Fund Grant.
\end{acknowledgments}


\end{document}